\documentclass[floatfix,aps,showpacs,superscriptaddress]{revtex4}
\usepackage{graphicx}

\begin{document}

\newcommand{\x}{{\bf x}}
\renewcommand{\k}{{\bf k}}
\def\mn#1{\marginpar{*\footnotesize \footnotesize #1}{*}}
\def\isum{{\sum\!\!\!\!\!\!\!\int}}
\def\bfzero{{\bf 0}}
\newcommand{\be}{\begin{equation}}
\newcommand{\ee}{\end{equation}}
\newcommand{\bea}{\begin{eqnarray}}
\newcommand{\eea}{\end{eqnarray}}
\newcommand{\bdm}{\begin{displaymath}}
\newcommand{\edm}{\end{displaymath}}
\newcommand{\bfk}{{\bf k}}
\def\hF{{\bar F}}
\newcommand{\bfp}{{\bf p}}
\newcommand{\ha}{{\hat a}}
\newcommand{\bfr}{{\bf x}}
\newcommand{\bfz}{{\bf 0}}
\newcommand{\veps}{\varepsilon}
\newcommand{\non}{\nonumber \\}
\def\lsim{\raise0.3ex\hbox{$<$\kern-0.75em\raise-1.1ex\hbox{$\sim$}}}
\def\gsim{\raise0.3ex\hbox{$>$\kern-0.75em\raise-1.1ex\hbox{$\sim$}}}

\title{Nonperturbative Effects on $T_c$ of Interacting Bose Gases in Power-Law Traps}
\author{O.~Zobay}
\affiliation{Institut f\"ur Angewandte Physik, Technische Universit\"at
Darmstadt, 64289 Darmstadt, Germany}
\author{G.~Metikas}
\affiliation{Department of Mathematics, Imperial College, London SW7 2AZ, United Kingdom}
\author{H.~Kleinert}
\affiliation{Institut f\"ur Theoretische Physik, Freie Universit\"at Berlin, Arnimallee 14, 14195 Berlin, Germany}
\begin{abstract}

The critical temperature $T_c$ of an interacting Bose gas trapped in a general power-law potential $V(\x)=\sum_i U_i|x_i|^{p_i}$ is calculated with the help of variational perturbation theory. It is shown that the interaction-induced shift in $T_c$ fulfills the relation $(T_c-T_c^0)/T_c^0= D_1(\eta)\hat a + D'(\eta)\hat a^{2\eta}+{\cal O}(\hat a^2)$ with $T_c^0$ the critical temperature of the trapped ideal gas, $\hat a$ the $s$-wave scattering length divided by the thermal wavelength at $T_c$, and $\eta=1/2+\sum_i p_i^{-1}$ the potential-shape parameter. The terms $D_1(\eta)\hat a$ and $D'(\eta)\hat a^{2\eta}$ describe the leading-order perturbative and nonperturbative contributions to the critical temperature, respectively. This result quantitatively shows how an increasingly inhomogeneous potential suppresses the influence of critical fluctuations. The appearance of the $\hat a^{2\eta}$ contribution is qualitatively explained in terms of the Ginzburg criterion.
\end{abstract}

\maketitle

\flushbottom

\section{\label{sec:intro}Introduction}

The realization of Bose-Einstein condensation (BEC) in dilute atomic vapors has renewed interest in the critical properties of weakly interacting Bose gases and, in particular,
their
transition temperature $T_c$. Important recent work
in this area concerns the role of the external trapping potential. For the homogeneous Bose gas, the shift in $T_c$ caused by $s$-wave contact interactions is in leading order completely due to long-wavelength, critical fluctuations that have to be described
nonperturbatively. It is now established that these fluctuations lead to a linear increase of the critical temperature with the $s$-wave
scattering length $a$, if the particle density is fixed \cite{BayBlaHol99,ArnMooTom01,ArnMoo01,KasProSvi01}.
A harmonic trapping potential, on the other hand,
suppresses the
critical long-wavelength fluctuations and reduces
 the fraction of atoms
taking part in nonperturbative physics
at the transition point.
 As a result,
the leading-order shift in $T_c$
can be calculated by simple pertubative methods,
for instance the
 mean-field (MF) approximation to the Landau-Pitaevskii
equation \cite{GioPitStr96,ArnTom01}. Interestingly, the shift
here is a {\it decreasing} linear function of the scattering length $a$.

Work on the critical temperature of Bose gases has so far been mainly concerned with homogeneous and harmonically trapped systems. As outlined above, in these two situations very different physical mechanisms determine the shift of $T_c$. This observation naturally motivates an
investigation of the crossover between the two cases.
In this paper we shall interpolate between these limits
by studying
the condensation in  a general power-law potential, whose parameters can be varied
continuously.
In this way we shall obtain a deeper
understanding of how the increasing inhomogeneity of the
potential suppresses the critical fluctuations
and changes
nonperturbative into
perturbative physics.
The power-law potentials under study are given by
\be\label{potential}
V({\bf x}) = \sum_{i=1}^3E_i \left|\frac {x_i} {L_i}\right|^{p_i}
\ee
with $E_i$ and $L_i$ denoting energy and length scales. If all powers $p_i$ are set equal to $2$, we recover the harmonic potential, whereas in the limit of all $p_i$ diverging, $V(\x)$ approaches a box shape characteristic of the homogeneous Bose gas.

First investigations of the crossover behavior of the critical temperature in these potentials were recently reported in
Refs.\ \cite{Zob04,ZobMetAlb04}.
In \cite{Zob04}, the shift in $T_c$ for these potentials was determined within mean-field theory in the thermodynamic limit. Extending
earlier first-order calculations \cite{ShiZhe97a,PI}, it was shown that up to second order in the scattering length $a$, the MF shift has
an expansion
\be\label{TcShift}
\frac{T_c-T_c^0}{T_c^0}=D_1^{(MF)}(\eta)\hat a+D'_{(MF)}(\eta)\hat a
^{2\eta}+D_2^{(MF)}(\eta)
\hat a^2+o(\hat a^2)
\ee
with $T_c^0$
the critical temperature of the noninteracting gas,
and
$\hat a\equiv a/\lambda_{T_c}$
the scattering length
measured in units of the thermal wavelength
$ \lambda _T\equiv  \sqrt{2\pi \hbar ^2/k_BTm}$
for particles of mass $m$
at a temperature $T$.
The exponent of the second term is twice the
{\em shape parameter\/} of the potential,
\be\label{eta}
\eta = \sum_{i=1}^3
\frac 1 {p_i}
 +\frac 1 2,
\ee
so that $2 \eta $
grows from  $1$
to
 $4$ as the shape
 changes from
 homogeneous to harmonic.
The coefficients  $D'_{(MF)}(\eta)$ and $D_{1,2}^{(MF)}(\eta)$ are respectively given explicitly or through simple quadratures. These results provided first insights into the crossover in the behavior of $T_c$ between homogeneous and inhomogeneous potentials. However, since mean-field theory does not account for critical fluctuations, it can only provide a
rough first estimate,
especially in the quasi-homogeneous regime, and needs
to be improved by more sophisticated methods.

One possible pathway for taking critical fluctuations into account was explored in Ref.\ \cite{ZobMetAlb04}, where the shift in $T_c$ was calculated with the help of a renormalization group (RG) method initially developed for studying the harmonically trapped gas \cite{MetZobAlb04}. The results were found to be in good qualitative agreement with mean-field theory.
As a main advantage,
the RG approach employed in that work
gave
a simple and transparent tool to
compute the critical temperature for a
wide
 range of potential shapes and interaction strengths.
A disadvantage was, however,
that the results were
 mainly numerical and rendered only
limited physical insight into the system. Furthermore, the
calculation required several unsystematic
approximations which are difficult to improve.

The purpose of this paper is to present a
more systematic approach to the problem by making use of
field-theoretic
 variational perturbation theory (VPT). VPT is a powerful
resummation method for divergent
perturbation series \cite{PI},
which has been extended to
quantum field theory and its anomalous
dimensions in Ref.~\cite{SCT,KS}.
It has led to a prediction
of the critical exponents of superfluid helium
with unprecedented accuracy
\cite{SCT2,KS}, as confirmed
for the exponent
 $\alpha$ of the specific heat of helium
by satellite experiments \cite{Lipa}.

In the context of BEC, field-theoretic VPT has been applied successfully
to determine
 the shift of the critical temperature of the homogeneous Bose gas
from a five-loop perturbation expansion
 \cite{Kle03}, extended recently
to six and seven loops in
 \cite{Kas03}.
In the present work we shall
describe the trapped Bose gas
in the thermodynamic limit, in which the
trap
is so wide that we may
apply,
as in \cite{Zob04,ZobMetAlb04}, the local-density approximation (LDA). The
system is
treated as locally homogeneous at any point $\bfr$ with an effective chemical potential $\mu_{\rm eff} = \mu-V(\bfr)$, where $\mu$ denotes the global chemical potential. In this way,
we can make contact with the high-order perturbative loop expansions that were derived in Refs.~\cite{Kle03,Kas03}
for classical three-dimensional
$\phi^4$-theories of {\it homogeneous} systems. Because of
dimensional reduction \cite{Zin89}, the effective classical theory
\cite{FK,PI} can be directly used
to describe critical properties of
the (quantum-mechanical) Bose gas below second order
in the scattering length \cite{BayBlaHol99,ArnMooTom01}.

In this work we shall combine the
high-order loop expansions with the LDA
to derive a perturbative expansion for
the particle number of the trapped system
in powers of $\ha=a/\lambda_T$.
From this expansion we extract, with the help of
field-theoretic VPT, the critical particle number and
the shift of $T_c$. The main results are:
(i) For small $\ha$, the shift of $T_c$ is shown to exactly follow a behavior $(T_c-T_c^0)/T_c^0= D_1(\eta)\hat a + D'(\eta)\hat a^{2\eta}+{\cal O}({\hat a}^2)$ in generalization of the mean-field result (\ref{TcShift}). The term proportional to $\ha^{2\eta}$ represents the leading nonperturbative effects, whereas $D_1(\eta)$ can be calculated perturbatively as discussed in \cite{Zob04,ShiZhe97a}. The second-order contribution, which we will not study in detail, contains terms proportional to $\hat a^2$ and $\hat a^2\ln \hat a$. (ii) We compute the coefficient $D'(\eta)$ for $\eta < 1$ using VPT, and in this way arrive at a quantitative description of the behavior of $T_c$ below second order in the scattering length. (iii) Following \cite{ArnTom01}, we give a qualitative explanation for the $\ha^{2\eta}$ behavior of $T_c$ based on the
Ginzburg criterion \cite{GI}.

The paper is organized as follows. In Sec.\ II, the physics of ideal
Bose gases in power-law potentials is
 briefly reviewed.
 In Sec.\ III we show with the help of general scaling
arguments why the critical temperature obeys a law of
 the form (\ref{TcShift}), and  give a physical
 interpretation for the appearance of the nonanalytic term.
Section IV contains the calculation of the nonperturbative
coefficient $D'(\eta)$, and illustrates the behavior of the critical temperature by means
of a numerical example. Summary and conclusions are given in Sec.\ V.

\section{\label{sec:trap}Ideal Bose gases in power-law traps}

In this section we summarize some known
properties of ideal Bose gases in power-law traps
that are needed later.
The notation follows Refs.\ \cite{ShiZhe97a} and \cite{BagPriKle87}. We consider a system of $N$ ideal bosons of mass $m$
trapped in the power-law potential (\ref{potential})
characterized by the shape parameter  $\eta$ introduced in Eq.~(\ref{eta}).
We define the characteristic volume by
\be\label{Vchar}
V_{{\rm char}}^{2(\eta+1)/3}= 8 \left(\frac{\hbar^2}m\right)
^{\eta-\frac 1 2}
\prod_{i=1}^3\frac {I(p_i)L_i}{E_i^{{1}/{p_i}}},
\ee
where $I(p_i)\equiv\Gamma(1/p_i)/p_i$
and $\Gamma(z)$ denotes the  usual
gamma function. The quantity $V_{{\rm char}}$ provides an estimate for
the volume occupied by the one-particle ground state in the trap.

For calculations in the local-density approximation
 it is useful to convert spatial integrations involving the trap
potential into
energy integrations
according to the rule
$\int d^3x\, f[V(\bfr)] = \int d\varepsilon \tilde\rho(\varepsilon) f(\varepsilon)$. In this way, we can easily deal with power-law potentials of any type. The
density $\tilde\rho(\varepsilon)$ is the area
of the equipotential surface $V(\bfr)=\veps$. As shown in \cite{Zob04,ZobMetAlb04}, $\tilde\rho(\varepsilon)$ is given by
\be\label{rho}
\tilde\rho(\varepsilon) = \frac {V_{{\rm char}}^{2(\eta+1)/3}} {\Gamma(\eta-1/2)}\left(\frac m{\hbar^2}\right)^{\eta-1/2}  \varepsilon^{\eta-3/2}.
\ee

As announced above, we
 shall treat
the thermodynamic limit, in which
 $N$ and $V_{{\rm char}}$ go to infinity
at fixed
$N/V_{{\rm char}}^{2(\eta+1)/3}$.
The equation of state for the ideal Bose gas above the condensation point is then given by \cite{LiCheChe99,Yan00}
\be\label{Nbose}
N =
\frac 1 {(2\pi)^{3/2}} \left(\frac m {\hbar^2\beta}\right)^{\eta+1}
V_{{\rm char}}^{2(\eta+1)/3} \zeta_{\eta+1}(z),
\ee
where
$\beta\equiv 1/k_BT$
denotes the inverse temperature,
$\mu$ the chemical potential, and $z=\exp(\beta\mu)$ the fugacity. The
Bose-Einstein
functions
$\zeta_{\lambda}(z)  \equiv   \sum_{k=1}^{\infty} z^k/k^{\lambda}$
are polylogarithmic functions \cite{ch7}. The spatial density distribution of the gas is determined
by the relation
\be \label{id_dens}
n({\bf x}) = \lambda_T^{-3}\zeta_{3/2}\left(e^{\beta[\mu- V({\bf x})]}\right).
\ee
The condition for Bose-Einstein condensation
$N=\int d^3x\, n({\bf x}, \mu = 0)$
can be obtained from Eq.\ (\ref{Nbose})
by setting $z=1$ or directly from Eq.\ (\ref{id_dens}) and reads \cite{BagPriKle87}
\be\label{Tc_id}
N = \frac 1 {(2\pi)^{3/2}} \left(\frac m {\hbar^2\beta_c^0}\right)^{\eta+1}
V_{{\rm char}}^{2(\eta+1)/3} \zeta(\eta+1),
\ee
where we have replaced
 $\zeta _ \nu (1)$ by
 Riemann's $\zeta$-function
$\zeta( \nu )
=\sum_{n=1}^\infty 1/n^ \nu $, and $\beta_c^0=1/k_BT_c^0$ denotes the inverse critical temperature of the ideal gas.

\section{\label{sec:general}General behavior of critical temperature}

We now turn to the interacting Bose gas where we assume,
 as usual, that the interaction
is effectively a delta-function potential
which is completely characterized
by the $s$-wave scattering
length $a$.
In
the local-density approximation
to the thermodynamic limit,
the trapped particle number
$N$ at given temperature $T$ and chemical potential $\mu$
 can be calculated
from the integral
\be\label{Nlda}
N(T,\mu) = \int d^3x\, n_{\rm tr}(\bfr; \mu, T) \approx \int d^3x\, n_{\rm hom}(\mu-V(\bfr), T),
\ee
where
 $n_{\rm tr}$ is the trapped particle density
and $n_{\rm hom}$ the density of the homogeneous gas. As we briefly explain at the end of this section, we expect the LDA to be applicable if the condition $\lambda_{T_c}^2/a \ll V_{\rm char}^{1/3}$ is fulfilled, where $\lambda_{T_c}$ denotes the thermal wavelength at the condensation point and $V_{\rm char}$ is defined in Eq.\ (\ref{Vchar}). Obviously, for a fixed $\lambda_{T_c}$ this condition can always be met by making the trap wide enough (and increasing the particle number accordingly).

In the following, we want to apply Eq.\ (\ref{Nlda}) to explain why the critical temperature follows a behavior as indicated in Sec.\ I. A more detailed calculation
will be presented in Sec.\ IV. Consider the perturbation
 expansion
of the homogeneous density $n_{\rm hom}$
in powers
 of $\hat a=a/\lambda_T$
\be\label{nexp}
n_{\rm hom}(T,\Delta\mu)=n_{\rm hom}^{(0)}(T,\Delta\mu) +
n_{\rm hom}^{(1)}(T,\Delta\mu)\hat a+
n_{\rm hom}^{(2)}(T,\Delta\mu)\hat a^2
+{\cal O}\left(\hat{a}^3\right),
\ee
 where
 $\Delta\mu\equiv \mu_c-\mu$  is the negative distance of the
chemical potential $\mu$ from its critical value $\mu_c$
at temperature $T$.
Our definition of $\Delta \mu$ ensures that
it is positive above the transition.

The omitted higher-order terms in the expansion (\ref{nexp})
depend on the details of the particle interactions.
Being interested in
contributions below second order, we can disregard these details and work only with a contact interaction.
Note that in Eq.\ (\ref{nexp}) we have
neglected logarithmic terms appearing at second and higher order
in $\hat a$. These terms enter via the critical chemical potential $\mu_c$ which contains a contribution proportional to $\hat a^2\ln \hat a$ \cite{ArnTom01}. As the calculation of Sec.\ IV shows, the omitted logarithmic terms are not relevant for determining the shift in $T_c$ below second order.

From Eq.\ (\ref{nexp}),
we can convince ourselves
that there must exist
a perturbative second-order contribution to the critical particle number. This justifies the inclusion of a term proportional to $\hat a^2$ in the general expression for the shift in $T_c$, as mentioned at the end of Sec.\ I.
Indeed, since the expansion (\ref{nexp})
becomes arbitrarily accurate when we go sufficiently far away
from the critical region, we can split the spatial
integral in Eq.\ (\ref{Nlda}) into a part near
 the trap center and a remainder:
\be
N(T,\mu_c) = \int_{V(\bfr)\le V_0} d^3x\, n_{\rm hom}(\mu_c-V(\bfr), T)+\int_{V(\bfr)> V_0} d^3x\, n_{\rm hom}(\mu_c-V(\bfr), T).
\label{@fterm}
\ee
Here, $V_0$ denotes an energy above which the perturbative expansion of the density becomes accurate.
Inserting the expansion (\ref{nexp}) into the second integral, we see that $N(T,\mu_c)$ contains a contribution which is of second order in the scattering length.
Since the expansion for $n_{\rm hom}$ contains a term proportional
to $\hat a^2\ln \hat a$ as mentioned above, there will also be such a contribution in the exact second-order result (compare to Ref.\ \cite{ArnTom01} for the harmonic case).

The expansion coefficients $n_{\rm hom}^{(0)}$ and $n_{\rm hom}^{(1)}$ in the perturbation series (\ref{nexp}) remain finite in the critical limit $\Delta \mu\to 0$. From Eq.\ (\ref{Nlda}) we thus obtain well-defined perturbative contributions to the critical particle number in zeroth and first order in $\hat a$. The zeroth-order contribution is just the critical particle number of the ideal gas. However, all other coefficients $n_{\rm hom}^{(i)}$, $i\ge 2$, are
infrared-divergent in the critical limit of $\Delta\mu\to 0$ where they behave like $ 1/ \sqrt{ \Delta \mu}^{i-1}$.
Nevertheless, as shown in Refs.\ \cite{Kle03,Kas03},
we can make use of resummation techniques to extract information about critical properties from the expansion. If
we focus on effects below second order
in the scattering length, it is sufficient to consider
only
the leading divergence $n_{\rm hom}^{(i,{\rm div})}$ at each order, i.e., $n_{\rm hom}^{(i)}=n_{\rm hom}^{(i,{\rm div})} + o((\Delta \mu)^{(i-1)/2})$.
This leads to the power series
\be\label{bi}
\Delta n_{\rm hom}^{\rm div} (T,\Delta\mu) \equiv \sum_{i=2}^\infty
n_{\rm hom}^{(i,{\rm div})}(T, \Delta \mu)
\hat a ^i
  =\hat a\sum_{i=2}^\infty  b_i
\,  \left(\frac {\hat a}{\sqrt{ \beta \Delta\mu}}\right)^{i-1}.
\ee
 Above the transition
where
the chemical potential $\mu$ is smaller than $\mu_c$
so that
 $\Delta\mu>0$, we insert
(\ref{bi}) into (\ref{Nlda}) and obtain the leading-order divergent contribution to the trapped particle number
\bea\label{Ndiv1}
\Delta N^{{\rm div}}(T,\Delta\mu)&= &\!\!\!\int d^3x\, \Delta n_{\rm hom}^{{\rm div}}(T,\Delta\mu+V({\bf x}))
=\!\hat a  \int d^3x\,\sum_{i=2}^\infty b_i
 \left(\frac {\hat a}{\sqrt{ \beta [\Delta\mu+V({\bf x})]}}\right)^{i-1}.
\eea
Converting the spatial integral into an energy integral
with the help of (\ref{rho})
 and performing this integral with analytic regularization,
we obtain
\bea\label{Ndiv}
\Delta N^{{\rm div}}(T,\Delta\mu)
& =&\!\!\! A \int d\veps\, \veps^{\eta-3/2}\hat a
\sum_{i=2}^\infty b_i \left(
\frac {\hat a}
{\sqrt{ \beta (\Delta\mu+\veps)}}\right)^{i-1}\non
&=&\! C (\Delta \mu)^{\eta} \sum_{i=2}^\infty
b_i \left( \frac {\hat a}
{\sqrt{ \beta \Delta\mu}}\right) ^i\frac{\Gamma(i/2-\eta)}
{\Gamma(i/2-1/2)}\non
&=&(\Delta \mu)^{\eta}\,h_1\!\!\left(\frac {\hat a}
{\sqrt{ \beta \Delta\mu}}\right),
\eea
where irrelevant constants have been absorbed into
the coefficients $A$ and $C$.
Note that the factors
 $\Gamma(i/2-\eta)$  in the
coefficients
cause divergences for $\eta\to 1$.
We ignore this issue for the moment and defer its discussion
to Sec.\ IV.
The main property
of
(\ref{Ndiv}) is that
 the function $h_1$ in the final expression
 depends only on the ratio $  {\hat a}/
{\sqrt{ \beta \Delta\mu}}$.
If the number of particles is to remain finite
 in the critical limit $ \Delta \mu\rightarrow 0$,
 the
limiting behavior of
 $h_1({\hat a}/
{\sqrt{ \beta \Delta\mu}})$
must be
$h_1(  {\hat a}/
{\sqrt{ \beta \Delta\mu}})\propto
( {\hat a}/
{\sqrt{ \beta \Delta\mu}})^{2\eta}$.
It follows that the critical $\Delta N^{{\rm div}}$ behaves like $\hat a^{2\eta}$.

Combining this result with the perturbative first-order contribution mentioned above, we find that the change $\Delta N_{\rm crit}$ in the critical particle number at fixed temperature is given by
\be\label{Nshift1}
\frac{\Delta N_{\rm crit}}{N_{\rm crit}^{0}} =
C_1(\eta) \hat a + C'(\eta)
\hat a^{2\eta}+ {\cal O}(\hat a^2)
\ee
with $N_{\rm crit}^{0}$ the critical particle number of the ideal gas, and $C_1(\eta)$, $C'(\eta)$ proportionality constants depending on the potential shape.
From this behavior we immediately deduce
the change of the critical temperature
at fixed particle number to behave like
\be\label{tcshift1}
\frac{\Delta T_c}{T_c^{0}} = D_1(\eta) \hat a +
D'(\eta) \hat a^{2\eta}+ {\cal O}(\hat a^2)
\ee
with coefficients $D_1(\eta)$ and $D'(\eta)$
which follow trivially from $C_1(\eta)$ and $C'(\eta)$ [compare with Eq.\ (\ref{tcshift}) below].

\begin{figure}
\begin{center}
\includegraphics[width=7cm]{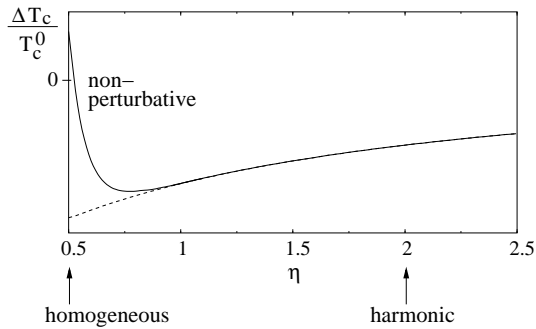}
\caption{Schematic diagram of shift in $T_c$
for a fixed small scattering length $\hat a$
as a function of
the potential shape parameter $\eta$. The upper curve shows the full
result below second order according to Eq.\ (\protect\ref{tcshift1}),
the dashed lower curve only displays the perturbative contribution linear in $\hat a$.
The nonperturbative contribution
decreases fast
with
 growing
inhomogeneity.
\label{fig1} }
\end{center}
\end{figure}

To discuss the physical contents of Eq.\ (\ref{tcshift1})
we anticipate some of the results of the next section, and schematically show in Fig.\ \ref{fig1} the behavior of the critical temperature at a fixed, small value of $\hat a$ as a function of the potential shape parameter $\eta$. The full curve shows the combined perturbative and nonperturbative contributions, whereas
the dashed curve displays only the perturbative (linear) term.
In the homogeneous limit we have $2\eta=1$, so that the shift
below second order in $\hat a$ is purely linear as we expect from earlier studies \cite{BayBlaHol99,ArnMooTom01,HolBayBla01}. In this case, both contributions [i.e., $D_1(\eta) \hat a$ and $D'(\eta)\hat a^{2\eta}$] are
 of comparable size at any value of $\hat a$.

The situation changes when we enter the
inhomogeneous regime where $2\eta > 1$. As displayed in Fig.\ \ref{fig1}, for sufficiently small, fixed $\hat a$  and growing $\eta$ the nonperturbative contribution $D'(\eta) \hat a^{2\eta}$ rapidly becomes very small compared to the perturbative term. This kind of behavior is independent of the detailed form of $D'(\eta)$ [note that in Fig.\ \ref{fig1}, we ignore the (unphysical) divergence of our approximation (\ref{tcshift1}) in a narrow vicinity of $\eta=1$; as discussed in Sec.\ IV, this is expected to be remedied in a higher-order expansion]. Equation (\ref{tcshift1})
 describes quantitatively how the growing inhomogeneity of the potential
reduces the influence
 of critical fluctuations on the transition temperature.

Following the arguments of Ref.\ \cite{ArnTom01}, we can also give a physical explanation for the
appearance of the $\hat a^{2\eta}$ term by estimating the
fraction of atoms actually taking part in nonperturbative effects. For
simplicity, consider the potential
$V(\bfr)=\sum_i V_0|x_i/r_0|^{\gamma}$.
Treating the  Bose gas above the transition point as classical, we find the mean-square width
\be
\langle x_i^2\rangle = r_0^2\frac{\Gamma(3/\gamma)}{(\beta V_0)^{2/\gamma}\Gamma(1/\gamma)},\quad i=1,2,3,
\ee
i.e., the cloud radius behaves like
\be
r_{\rm cloud}\sim\frac{r_0}{(\beta V_0)^{1/\gamma}}.
\ee
From the Ginzburg criterion
\cite{GI} it follows that nonperturbative effects only arise at (local)
chemical potentials $\mu_{\rm eff}$, for which
\be
\mu_c-\mu_{\rm eff} ~\lsim~ \frac{\hbar^2a^2}{m\lambda_T^4}.
\ee
Invoking the local density approximation, this means that the nonperturbative region around the
trap center has a radius of about
\be\label{rnp}
r_{\rm np}\sim r_0\left(\frac{\hbar^2a^2}{m\lambda_T^4V_0}\right)^{1/\gamma}.
\ee
The fraction of atoms within this nonperturbative spatial region is given by
\be
\left(\frac {r_{\rm np}}{r_{\rm cloud}}\right)^3 \sim \left(\frac a{\lambda_T}\right)^{6/\gamma}.
\ee
However, not all atoms within this region
actually take part in nonperturbative physics \cite{ArnTom01}.
From the homogeneous system we infer that only a fraction $a/\lambda_T$ are actually involved in these effects. For the trap, this means that the fraction of ``nonperturbative" atoms scales like
\be
\frac a{\lambda_T} \left(\frac{r_{\rm np}}{r_{\rm cloud}}\right)^3\sim \left(\frac a{\lambda_T}\right)^{(6/\gamma)+1}=\left(\frac a{\lambda_T}\right)^{2\eta},
\ee
which explains the appearance of the nonanalytic term in Eq.\ (\ref{Nshift1}).

At this point, it is also convenient to explain the estimate for the validity of the LDA given above, again following the arguments of Ref.\ \cite{ArnTom01}. Nonperturbative effects involve fluctuations with wavelengths $\lambda_T^2/a$ and larger. For the LDA to be applicable, the size $r_{\rm np}$ of the nonperturbative region around the trap center should thus be much larger than $\lambda_T^2/a$. Since Eq.\ (\ref{rnp}) can be rewritten as
\bdm
r_{np}\sim \left(\frac a{\lambda_T}\right)^{2/\gamma}V_{\rm char}^{(2/\gamma +1)/3}
\edm
by using Eq.\ (\ref{Vchar}), this condition immediately implies that $\lambda_T^2/a \ll V_{\rm char}^{1/3}$. In other words, the extension of the ground state has to be much larger than the minimum length scale $\lambda_T^2/a$ for critical fluctuations. With the help of Eq.\ (\ref{Tc_id}), we can also rephrase this statement as follows. With condensation taking place at temperature $T$, the trap has to be sufficiently wide, so that the critical atom number fulfills the condition
\bdm
a N^{1/(2\eta+2)}\gg \lambda_T.
\edm

\section{\label{sec:calc}Calculation of critical temperature}

After the general discussion of Sec.\ III,
we now turn to the actual calculation of the nonperturbative
coefficient $D'(\eta)$. Our result will be approximate
in two ways: first, we use some approximations to calculate the coefficients of the weak-coupling expansion for the trapped particle number.
Second,  the number of terms is limited to seven,
and the evaluation via
VPT leaves an error.
However, due to the stability and the typically exponentially fast convergence
 of VPT,
our results should
provide a satisfactory representation of the true behavior.

To simplify the notation for the following calculations,
we introduce the reduced homogenous density function
\be\label{red}
f(\beta(\mu_c-\mu))\equiv \lambda_T^3\, n_{\rm hom}(\beta\mu)
=\lambda_T^3 \,n_{\rm hom}(\beta\mu_c-\beta(\mu_c-\mu)).
\ee
This quantity implicitly depends, of course,
on the reduced scattering length $\hat a=a/ \lambda _T$, as indicated
in the expansion (\ref{nexp}).

Under the assumption of LDA,
$f(\Delta+\beta V(\bfr))$ with $\Delta=\beta\Delta\mu \ge0$
equals the reduced particle density
of the
trapped Bose gas at point ${\bf x}$.
For $ \Delta =0$, it describes
 the critical
density profile in the trap.
We now define the integral
\be\label{I}
I_\eta(\Delta) \equiv \int_0^{\infty} dv\,v^{\eta-3/2} f(\Delta+v).
\ee
Using Eqs.\ (\ref{rho}) and (\ref{red}), this can be recognized as a rescaled version of Eq.\ (\ref{Nlda}).
The shift in the critical temperature for a fixed particle number is then given by \cite{Zob04,ZobMetAlb04}
\be\label{tcshift}
\left(\frac{T_c}{T_c^{(0)}}\right)^{\eta+1} = \frac{\zeta(\eta+1)}{I_\eta(0)/\Gamma(\eta-1/2)}.
\ee
As sketched in the previous section, we shall first derive a
perturbative expansion for
$f(\beta(\mu_c-\mu))$ in powers of $\hat a$. For $\Delta\to 0$, the zeroth- and first-order terms of this expansion remain finite, whereas the
higher-orders terms suffer from infrared divergences.
The zeroth and first order can thus be directly inserted into Eq.\ (\ref{I}) and their contribution read off at $\Delta=0$. For the higher-order terms, we focus only on the leading-order divergence. This leads to an expansion in terms of $\hat a/\sqrt{\Delta+v}$. Inserting this result into (\ref{I}) and performing the integration over $v$, we obtain the weak-coupling expansion for $I_\eta(\Delta)$
[compare with Eq.\ (\ref{Ndiv})]. This expansion is finally resummed to find the coefficient $D'(\eta)$.

Let us temporarily return to the more familiar unscaled quantities to outline further details of the calculation. Using the LDA and Eq.\ (\ref{rho}), the trapped particle number is given by
\bea\label{nlda}
N(T,\mu)&=&\int d^3x\, n_{\rm hom}(T,\mu-V(\bfr))\non
&=&\left(\frac m{\hbar^2}\right)^{\eta-1/2} \frac 1
{\Gamma(\eta-1/2)} V_{\rm char}^{2(\eta+1)/3}
\int d\veps\, \varepsilon^{\eta-3/2}n_{\rm hom}(T,\mu-\veps).
\eea
We now express
$n_{\rm hom}(T,\mu-\veps)$ in terms of the Green function
of the interacting homogeneous system:
\bdm
G_{\rm hom}(\k,\omega_n;\mu, T)=\frac 1{i\omega_n-[\veps_{\k}-\mu+\hbar\Sigma(\k,\omega_n;\mu,T)]/\hbar},
\edm
where
 $\veps_{\k}=\hbar^2 \k^2/2m$ with momentum $\k$,
$\omega_n=2\pi n/\beta\hbar$ with integer $n$ are the Matsubara frequencies, and $\Sigma
(\k,\omega_n;\mu ,T)$ is the proper self energy. 
This leads to
\bea\label{nlda2}
N(T,\mu)
&=&-\left(\frac m{\hbar^2}\right)^{\eta-1/2} \frac {V_{\rm char}^{2(\eta+1)/3}\hbar^3}
{\Gamma(\eta-1/2)\beta\hbar(2\pi)^3}\non
&&\times \int d\veps\, \varepsilon^{\eta-3/2}\isum_k \frac 1{i\omega_n-(\veps_{\k}-\mu+\veps)/\hbar -\Sigma(\k,\omega_n;\mu-\veps,T)}.
\eea
The symbol $\sum\!\!\!\!\!\int_{\ k}$ with $k=(\k,\omega_n)$ denotes integration and summation over
all momenta and Matsubara frequencies. The last term  in
(\ref{nlda2}) is conveniently rearranged to
\bea\label{nlda3}
\isum_k \frac 1{i\omega_n-[\veps_{\k}-(\mu-\veps)+\hbar\Sigma({\bf 0},\omega_n;\mu-\veps,T)]/\hbar -[\Sigma(\k,\omega_n;\mu-\veps,T)-\Sigma({\bf 0},\omega_n;\mu-\veps,T)]}.
\eea
In this expression, we
 use the ``mass-renormalized" Green function
\bdm
G_0(\k,\omega_n;\mu,T)=\frac 1{i\omega_n-[\veps_{\k}-\mu+\hbar\Sigma({\bf 0},\omega_n;\mu,T)]/\hbar}
\edm
as the
 free Green function for a
 perturbative expansion. In this way,
we obtain
 two different contributions
to the trapped particle number $N(T,\mu)=N^{(1)}(T,\mu)+N^{(2)}(T,\mu)$,
namely,

(i) the zero-order term
\be\label{N1}
N^{(1)}(T,\mu)=C\int d\veps\, \varepsilon^{\eta-3/2}\isum_k \frac 1{i\omega_n-[\veps_{\k}-(\mu-\veps)+\hbar\Sigma({\bf 0},\omega_n;\mu-\veps,T)]/\hbar}
\ee
with $C$ the prefactor in front of the integral of Eq.\ (\ref{nlda2}) and

\begin{figure}
\begin{center}
\includegraphics[height=7cm,angle=270]{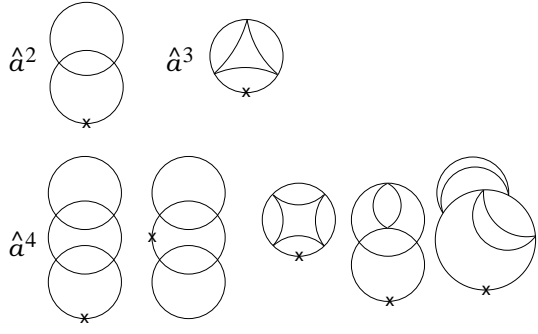}
\caption{Diagrams contributing to $N^{(2)}$ up to order $\hat a^4$. The crosses denote the joining of two Green functions.
\label{fig2} }
\end{center}
\end{figure}

(ii) higher-order
contributions
pictured by  the Feynman
diagrams in Fig.\ \ref{fig2}
up to five loops. Using the zero Matsubara frequency contributions of these diagrams, we will calculate a perturbation series for the leading infrared-divergent contribution $n_{\rm hom}^{(2,\rm div)}(T,\mu)$ to the homogeneous density.
From this we obtain the second contribution $N^{(2,\rm div)}(T,\mu)=\int d^3x\, n_{\rm hom}^{(2,\rm div)}(T,\mu-V(\bfr))$ to the trapped particle number. We note that the nonzero Matsubara frequency modes are expected to contribute only in second order in $\hat a$ to the critical particle number
\cite{BayBlaHol99}.

\begin{figure}
\begin{center}
\includegraphics[height=7cm,angle=270]{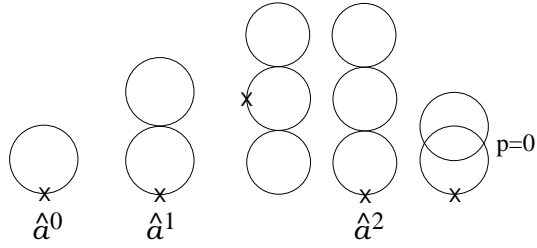}
\caption{Diagrams contributing to $N^{(1)}(T,\mu)$ up to order $\hat a^2$. The sunset subdiagram is taken at the external momentum $p=0$.
\label{fig3} }
\end{center}
\end{figure}

First we discuss $N^{(1)}(T,\mu)$. The
relevant diagrams for the perturbative evaluation of Eq.\ (\ref{N1}) 
are shown in Fig.\ \ref{fig3} up to three loops. As in Fig.\ \ref{fig2}, an $n$-loop diagram contributes to order $\hat a^{n-1}$ to the perturbation series. The contributions of zeroth and first order in
$\hat a$ are very easily calculated.
Since they are convergent in the limit of $\Delta\to 0$,
we find
up to first order in $\hat a$:
\bea\label{pert_exp}
I_\eta(0) &=& \Gamma(\eta-1/2)\zeta(\eta+1)
+\hat a \int_0^\infty dv\, v^{\eta-3/2} 4F_{1/2}(v)[\zeta(3/2)-F_{3/2}(v)]+\dots\non
&=&\Gamma(\eta-1/2)\zeta(\eta+1) +\Gamma(\eta-1/2)C_1(\eta)\hat a+\dots
\eea
This expression is equivalent to the first-order mean-field results
 of Refs.\ \cite{Zob04,ShiZhe97a}. The
 perturbative contribution is nonzero
in the homogeneous limit $\eta=1/2$ \cite{Kle03,Zob04}.

The higher-order diagrams are divergent in the limit of $\Delta\to 0$. We shall
not use the whole set of diagrams for our calculation, but restrict ourselves for simplicity to the Hartree-Fock (HF) approximation in which we take only diagrams into account that consist purely of simple bubbles (such as the first four in Fig.\ \ref{fig3}). Unfortunately, it seems difficult to estimate the consequences of this approximation or to even go beyond it, but we expect that it captures the main features of the actual behavior.
At any rate, the HF approximation is interesting in itself since the resummation can be done exactly and provides a nice illustration of our approach.
We also find that it leads to the result already obtained in Ref. \cite{Zob04} by a very different derivation. It should be remarked that our calculation shows that the HF approximation, which is equivalent to the so-called `mean-field' description (see, e.g., \cite{Zob04}), already includes nonperturbative effects.

In general, the HF approximation for a contact interaction consists of writing the proper self energy as
\be
\Sigma(\bfk,\omega_n;\mu,T)=-\frac {2g}{\hbar^2\beta}\frac1{(2\pi)^3}\isum_{k'} G(\bfk',\omega_{n'};\mu,T)
\ee
with $g=4\pi\hbar^2 a/m$ \cite{FetWal71}. Inserting this expression
into Dyson's equation for the Green function
and
iterating the procedure
we see that the Hartree-Fock approximation is equivalent to summing over
all pure-bubble diagrams as mentioned above.
Alternatively,
we shall work out the bubble series directly. The starting point
is the mean-field equation for the homogeneous density
\be
\lambda_T^3 n_{\rm hom} = F_{3/2}(-\beta\mu+2\beta gn_{\rm hom})
\ee
where  $F_ \nu (x) = \zeta_ \nu (e^{-x})$, which is equivalent to the HF theory. With our scaled quantities, this equation reads
\be
f_{HF}(\Delta+v) = F_{3/2}(\Delta+v-\beta\mu_c+4\hat a f_{HF}) =
F_{3/2}\Big(\Delta+v+4\hat a\big[f_{HF}( \Delta +v)-\zeta(3/2)\big]\Big)
\ee
where we have inserted the lowest-order
equation
$\beta\mu_c=4\hat a\zeta(3/2)$
valid for small $\hat a$ \cite{ArnTom01}.
 The right-hand side
is now expanded in powers of $\hat a$:
\be\label{f_exp}
f_{HF}(\Delta+v)=\sum_{n=0}^{\infty}\frac{(-1)^n}{n!}
F_{3/2-n}(\Delta+v)(4\hat a)^n[f_{HF}(\Delta+v)-\zeta(3/2)]^n.
\ee
Solving this implicit equation  by iteration yields
\bea
f_{HF}(\Delta+v)&=&F_{3/2}(\Delta+v) - 4\hat a\,F_{1/2}(\Delta+v)\,\hF_{3/2}(\Delta+v) \nonumber\\
 && +(4\hat a)^2\,\left[ F_{1/2}^2(\Delta+v)\,\hF_{3/2}(\Delta+v) +
     \frac{F_{-1/2}(\Delta+v)\,
        \hF_{3/2}^2(\Delta+v)}{2} \right] \nonumber\\
&&+(4\hat a)^3\,\left[ - F_{1/2}^3(\Delta+v)\,
        \hF_{3/2}(\Delta+v)  -
     \frac{3\,F_{-1/2}(\Delta+v)\,
        F_{1/2}(\Delta+v)\,\hF_{3/2}^2(\Delta+v)}{2}\right.\nonumber\\
&&~~~~~~~~- \left.  \frac{F_{-1/2}(\Delta+v)\,
        {\hF_{3/2}^3(\Delta+v)}}{6} \right]+{\cal O}(\hat a^4) \label{hf_exp}
\eea
where the subtracted function $\hF_{ \nu }(x)
\equiv F_{ \nu }(x)-\zeta( \nu )$
 vanishes at $x=0$ for $ \nu >1$.
The individual terms in (\ref{hf_exp})
correspond to the HF-diagrams in the perturbation
expansion for the homogeneous density (with the nonzero Matsubara frequencies taken into account). The first two terms once more give
the previous expansion (\ref{pert_exp}).
Using the Robinson expansion for $ \zeta _ \nu (x)$ \cite{Rob51,PI}:
\begin{equation}
\zeta_{ \nu  }(e^{-x})
=  \Gamma (1- \nu )x^{ \nu -1}
+
\sum_{k=0}^\infty
 \frac{1}{k!}(-x)^k  \zeta ( \nu -k),
\label{@newzetaex}\end{equation}
we find that the terms of order $n\ge 2$
in (\ref{hf_exp})
diverge like $\hat a( \hat a/\sqrt{\Delta+v})^{n-1}$ in the limit
 $(\Delta +v)\rightarrow 0$, due to the leading
Robinson term  $\Gamma (1- \nu )x^{ \nu -1}$.
Focusing attention upon this
leading divergence, we
 replace each function $F_ \nu (x)$ by
its Robinson term. This step corresponds to dropping the nonzero Matsubara frequency contributions from the diagrams of Fig.\ \ref{fig3}.
With the help of a
computer algebra program, we obtain in this way from Eq.\ (\ref{hf_exp})
the following series expansion of the divergent part of the density
\be
f^{({\rm div})}_{HF}(\Delta+v) = -\frac{\pi^{3/2}(4\hat a )^2}{\sqrt{\Delta+v}}+ \frac{\pi^{5/2}(4\hat a )^4}{4(\Delta+v)^{3/2}}- \frac{\pi^{7/2}(4\hat a )^6}{8(\Delta+v)^{5/2}} +\frac{5\pi^{9/2}(4\hat a )^8}{64(\Delta+v)^{7/2}}+\dots\ .
\ee
Inserting this expression into
 Eq.\ (\ref I), we can easily perform
the integration over $v$
with the help of dimensional regularization. Note that it is crucial to carry out this step for a finite $\Delta > 0$, since otherwise all integrals would vanish according to Veltman's rule $\int d \veps \veps^ \alpha =0$ for all $\alpha$ \cite{PI,KS}.
The result of this calculation is
\be
I^{({\rm div})}_{\eta;HF}(\Delta) = \Gamma(\eta-1/2)\Gamma(-\eta)\Delta^{\eta} \sum_{n\ge 1} \frac{(\eta-n+1)\cdots(\eta-1)\eta}{n!}\left(\frac{16\pi \hat a^2}\Delta\right)^n,
\ee
which is easily recognized as the
series expansion of $\Gamma(\eta-1/2)\Gamma(-\eta)[(16\pi \hat a^2+\Delta)^\eta-\Delta^\eta]$.
In the critical limit
 $\Delta\to 0$, this becomes
\be\label{IdivHF}
I^{({\rm div})}_{\eta;HF}( 0 ) =
(16\pi)^\eta\Gamma(\eta-1/2)\Gamma(-\eta)\hat a^{2\eta}.
\ee
We see that due to the resummation procedure all diagrams in the perturbation series effectively contribute to the leading-order nonperturbative shift of the critical particle number.

The same result (\ref{IdivHF}) was also found in \cite{Zob04} using a
completely
 different approach. There, it was
pointed out that this contribution can be
derived from the behavior of
the critical trapped
density within a
region
around the trap center where
$0\le \beta V(\x) \,\lsim\, \hat a^2$.
This is the nonperturbative regime by
the Ginzburg criterion  \cite{GI}. Here, in contrast, we use the perturbation expansion, which is valid far away from the trap center, to obtain the same result.

We now turn to evaluating $N^{(2)}(T,\mu)$, i.e., the second contribution to the trapped particle number which is determined by the diagrams displayed in Fig.\ \ref{fig2}. For the calculation, we shall make use of the high-order perturbative loop expansions that were derived in Refs.\ \cite{Kle03,Kas03}. These expansions allow us to evaluate the most divergent contributions $n_{\rm hom}^{(2,{\rm div})}$ to the homogeneous density, as defined above. Proceeding along the lines indicated in the study of $N^{(1)}$, we use $n_{\rm hom}^{(2,{\rm div})}$ to first obtain a series expansion for $N^{(2,{\rm div})}$ and then to resum the series to find the change in the critical particle number. This time the resummation cannot be performed exactly, which leads us to apply variational perturbation theory (VPT).

A problem in this procedure concerns the fact that Refs.\ \cite{Kle03,Kas03} do not calculate $n_{\rm hom}^{(2,{\rm div})}$ in terms of $\mu$ as we need it, but rather as a function of  $\xi=-\beta[\mu-\hbar\Sigma(\bfzero,\mu)]$ (in other words, we would need the diagrammatic expansion leading to Fig.\ \ref{fig2} to be carried out similar to Fig.\ \ref{fig3}, i.e., including bubble contributions). In the following we will ignore this difference and approximate the exact expression
\be
N^{(2,{\rm div})}(\Delta=\beta(\mu_c-\mu))= {\rm const.}\times \int_0^\infty d\veps \,\veps^{\eta-3/2}
\lambda_T^3 n_{\rm hom}^{(2,{\rm div})}(\beta\mu_c-\Delta-\beta\veps)
\ee
by
\be\label{n21}
N^{(2,{\rm div})}(\Delta)\approx {\rm const.}\times \int_0^\infty d\xi' \,\xi'^{\eta-3/2}
\lambda_T^3 \widetilde n_{\rm hom}^{(2,{\rm div})}(\Delta+\xi').
\ee
In this equation $\widetilde n_{\rm hom}^{(2,{\rm div})}$ denotes the divergent homogeneous density as a function of $\xi$, i.e., $n_{\rm hom}^{(2,{\rm div})}(\beta\mu)=\widetilde n_{\rm hom}^{(2,{\rm div})}(\xi(\beta\mu))$. In Refs.\ \cite{Kle03,Kas03}, $\widetilde n_{\rm hom}^{(2,{\rm div})}(\xi)$ is calculated in terms of a high-order perturbative loop expansion:
\be\label{s2}
\lambda_T^3 \widetilde n_{\rm hom}^{(2,{\rm div})}(\xi)
=\hat a
\sum_{i=3}^\infty b_i\left(\frac{\hat a}{\sqrt{\xi}}\right)^{i-2}.
\ee
From power counting, we expect
 $\beta\hbar\Sigma(\bfzero,\Delta)$ to have an
expansion of the form $\hat a^2\sum_l s_l \,(\hat a/\sqrt{  \Delta})^{l-2}$ for the leading-order divergence. Using this expansion we see that the exact expression for $N^{(2)}$ and our approximation differ somewhat in the higher-order coefficients of the weak-coupling expansion. Since the results of the resummation are most strongly affected by the lower-order coefficients, we neglect the error introduced in this way. We expect that
this simplification will not change the main features of our results.

We thus insert the expansion (\ref{s2}) into
Eq.\ (\ref{n21}). The coefficients $b_l$ appearing in (\ref{s2}) are related to the $a_l$'s of
Ref.\ \cite{Kas03} by $b_l=384\pi^3(24\pi)^{l-2}a_l$.
Applying dimensional regularization, we now perform the integration
\bea\label{weak1}
\frac 1{\Gamma(\eta-1/2)}\int_0^{\infty}
dv\, v^{\eta-3/2} \lambda_T^3
n_{\rm hom}^{(2,{\rm div})}( \Delta +v) &=&
\frac {\hat a}{\Gamma(\eta-1/2)}\sum_{i=1}^{\infty}b_i\int_0^{\infty}
dv\,v^{\eta-3/2}
\left(\frac {\hat a}{\sqrt{\Delta+v}}\right)^{i-2}
\nonumber\\
&=&\Delta^\eta \sum_{i=1}^{\infty}b_i
\left(\frac{\hat a}{\sqrt{\Delta}}\right)^{i-1} \frac{\Gamma(i/2-1/2-\eta)}{\Gamma(i/2-1)}.
\eea
This result constitutes a weak-coupling expansion of $ N^{(2,\rm div)}(T,\mu)$. However, we have to consider the limit $ \Delta \rightarrow 0$ which is a strong-coupling limit. We
assume that the
integrand
in (\ref{weak1})
has a strong-coupling expansion \cite{PI,KS}
\be\label{strong1}
\frac 1{\Gamma(\eta-1/2)}\int_0^{\infty} dv\,v^{\eta-3/2} \lambda_T^3
n_{\rm hom}^{(2,{\rm div})}( \Delta +v)
= {\hat a}^{2\eta}\sum_{m=0}^{\infty}B_m\left(\frac {\hat a} {\sqrt{\Delta}}\right)^{-\omega' m}.
\ee
The leading power of $\hat a$
is fixed by
dimensional considerations, i.e.,
by
 formally equating (\ref{weak1}) with (\ref{strong1}),
since the result
can only depend on the parameter $\hat a/\sqrt{\Delta}$ [also see the discussion following Eq.\ (\ref{Ndiv})].
The subleading powers
are multiples of the
universal Wegner exponent
governing the approach to scaling
as shown in  \cite{Kle03,Kas03}. This exponent reflects the influence from the anomalous dimensions of quantum field theory.

Our task is now to determine the coefficient $B_0$ of
the strong-coupling expansion.
This will yield the leading-order contribution $N^{(2,{\rm div})}$
to the shift in $N$. The result is obtained from field-theoretic
VPT in analogy to the calculations
for the homogeneous system in Ref.~\cite{Kle03}.
As in that reference we
perform two variants of the calculation,
one by fixing $\omega'$ to have the known value
$ 0.805$, and one by determining $ \omega '$
order by order self-consistently.
At first glance, this approach still leaves grounds
for scepticism, since
the perturbation expansion obtained from the LDA is divergent in two ways.
First, the expansion coefficients grow
factorially, so that the radius of convergence is zero.
Second, the series has to be evaluated
at a very large argument
which goes to infinity when approaching
the critical point.
Fortunately, these two unpleasant properties
are familiar from
perturbation expansions
of critical exponents,
for which it has been shown that
a resummation by field-theoretic VPT leads
to the correct results.
We can thus also safely use this method
here.

\begin{figure}
\begin{center}
\includegraphics[width=7cm]{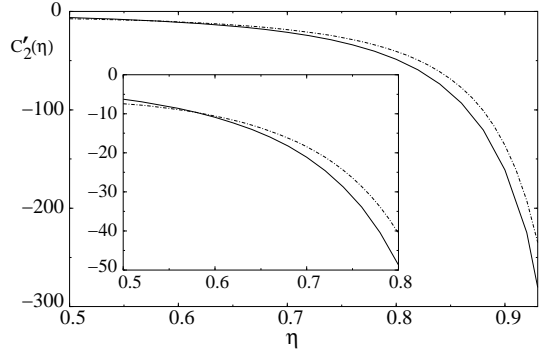}
\caption{Results
for $C'_2(\eta)$ from VPT:
(bold curve) fixed $\omega=0.805$; (dash-dotted) self-consistent determination of $\omega'$.  Note the divergence as $\eta\to 1$.
\label{fig4} }
\end{center}
\end{figure}

Within our approach, the change in the critical particle number is finally
obtained as an expansion
\be\label{Iresult}
\frac 1{\Gamma(\eta-1/2)}I_\eta(0) = \zeta(\eta+1)+C_1(\eta){\hat a} +[C'_{1}(\eta) +C'_{2}(\eta)]{\hat a}^{2\eta}+{\cal O}({\hat a}^2)
\ee
with $C_1(\eta)$ determined from Eq.\ (\ref{pert_exp}) and $C'_{1}(\eta)=(16\pi)^{\eta}\Gamma(-\eta)$
as in Eq.~(\ref{IdivHF}).
The coefficient
  $C'_{2}(\eta)$ emerges from
VPT. In Fig.\ \ref{fig4} we show the VPT calculation for $C'_{2}(\eta)$ using the seven-loop data from Ref.\ \cite{Kas03}. The result with self-consistent determination of $\omega'$ (bold curve) is compared to the calculation with a fixed value of $\omega'=0.805$. We see that both curves agree reasonably well; the calculation of $C'_{2}(\eta)$ thus does not depend too sensitively on the exponent. For $\eta\gsim 0.94$, the self-consistent calculation of $\omega'$ does not converge anymore; one would probably have to extend the resummation to higher orders to resolve this issue. More importantly, however, we see that the results for $C'_{2}(\eta)$ diverge in the limit of $\eta\to 1$. This behavior can be traced back to the presence of the $\Gamma$ functions in Eq.\ (\ref{weak1}). The same divergence also appears in the coefficient $C'_{1}(\eta)$. We will discuss the significance of these divergences below in connection with the calculation of the critical temperature.

\begin{figure}
\begin{center}
\includegraphics[width=7cm]{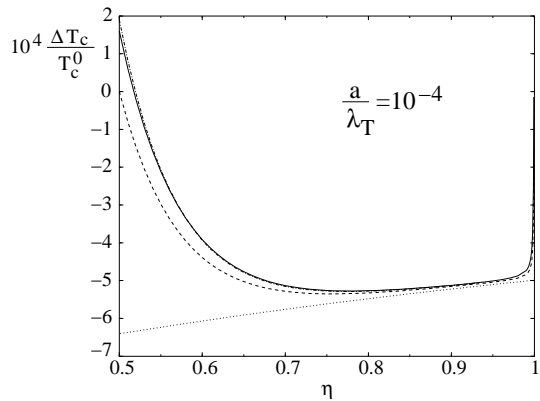}
\caption{Shift of critical temperature as a function of the potential shape parameter $\eta$ for fixed $a/\lambda_T=10^{-4}$: (full curve) Result including all terms in (\protect\ref{Iresult}) and $C'_2(\eta)$ calculated with fixed $\omega'=0.805$; (dash-dotted) same as full curve but with $\omega'$ determined self-consistently; (dashed) $C'_2(\eta)$ omitted (i.e., ``mean-field result"); (dotted) $C'_1(\eta)$ and $C'_2(\eta)$ omitted (i.e., only perturbative term).
\label{fig5} }
\end{center}
\end{figure}

In Fig.\ \ref{fig5}, we plot the final result,
i.e., the shift in the critical temperature, as determined from Eqs.\ (\ref{tcshift}) and (\ref{Iresult}). The figure reflects the characteristic qualitative features
discussed in Sec.\ III and confirms our previous conclusions. The shift is displayed as a function of
$\eta$ for a fixed value of $\ha=a/\lambda_T=10^{-4}$. The full and the dash-dotted curves show the complete result including all terms from Eq.\ (\ref{Iresult}). The two curves are respectively based on the calculation of $C'_2(\eta)$ with $\omega'$ kept fixed or determined self-consistently.
The discrepancy between these results can be considered an estimate for the error in the calculation of $T_c$ that is introduced by the resummation procedure. Since the curves are almost indistinguishable, we can conclude that the resummation contributes only a small inaccuracy in addition to potential further errors introduced by the other approximations.

For the dashed curve in Fig.\ \ref{fig5}, the term $C'_2(\eta)
\hat a^{2\eta}$ has been omitted,
it thus displays the mean-field result below second order \cite{Zob04}.
The dotted curve shows the perturbative contribution
due to $C_1(\eta)\hat a$.
We see that the full and the mean-field result closely approach the perturbative first-order approximation 
long before values of $\eta$ around 1 are reached. The divergence around $\eta=1$ introduced by the behavior of $C_1'$ and $C_2'$ is restricted to a very small interval. Thus it is reasonable to expect that the true behavior of $T_c$ in this small regime remains well described by the first-order approximation and that the divergence is only an artifact of our calculation. Furthermore, 
in Ref.\ \cite{Zob04} it was shown that within mean-field theory this divergence is compensated for by a pole in the second-order contribution to the critical particle number. We can also expect the same behavior in the present case, i.e., beyond mean-field theory.

\section{Summary and conclusions}

We have
calculated
the shift of the critical temperature of interacting Bose gases trapped in a
general
power-law potential
of the type $\sum_i U_i|x_i|^{p_i}$.
The objective was to understand how
this shift changes when we
 pass
from homogeneous
to harmonically trapped systems,
 interpolating between these limits
by changing the power of the potential.
While the homogeneous
case is influenced
strongly by nonperturbative critical fluctuations,
the harmonic case can be calculated  perturbatively.

We have restricted our attention
 to
the thermodynamic limit, which allowed us to
use the local-density approximation
in which the Bose gas is assumed to be locally homogeneous at each point. In Sec.\ III, we gave justification for this procedure.
The main result is the
shift formula
\be
\frac{T_c-T_c^0}{T_c^0}=D_1(\eta)\frac a{\lambda_T}+D'(\eta)\left(\frac a{\lambda_T}\right)^{2\eta}+{\cal O}(a^2).
\ee
It contains a linear, perturbative part,
 which is relevant for all potentials,
and a nonperturbative contribution proportional to $(a/\lambda_T)^{2\eta}$.
For small $a/\lambda_T$, the
 latter contributes significantly
only in the quasihomogeneous regime.

The presence of the $(a/\lambda_T)^{2\eta}$ term
was derived from scaling and resummation arguments,
and we gave a simple physical explanation for its appearance
based on the
Ginzburg criterion  \cite{GI}.
Our results show how the growing inhomogeneity
of the potential reduces the significance of critical fluctuations.

We also performed an explicit calculation of the
nonperturbative coefficient $D'(\eta)$.
In spite of the approximate character of the calculation,
the result
should be quite accurate.
Our approach was based on the resummation of divergent perturbation series with the help of
field-theoretic
variational perturbation theory. In the course of the derivation, it was shown that
 the simple Hartree-Fock
approximation also contains a nonperturbative contribution.

Finally, our study indicates
 that the higher-order $T_c$ shift in traps
with $\eta=1$,
for instance in $\sum_i U|x_i|^6$ potentials,
should be particularly interesting and difficult to investigate.
In the limit $\eta \rightarrow 1$,
we find a divergence in
the coefficient $D'(\eta)$
which governs
the nonperturbative contribution proportional
to $(a/\lambda_T)^2$.
It remains to be investigated
whether
this divergence is canceled
by other
genuinely second-order terms, as it is expected from mean-field theory. For future work, it might also be of interest to study the influence of other types of external potentials on the transition temperature, such as optical lattices \cite{KleSchPel04}.

\begin{acknowledgments}
O.Z.\ and G.M.\ thank
the Deutsche Forschungsgemeinschaft for financial support
of their stays at the FU Berlin through
the priority program SPP 1116 and the Forschergruppe ``Quantengase."
They also gratefully acknowledge stimulating discussions with B.\ Kastening, F.\ Nogueira, A.\ Pelster, and G.\ Alber.
Further partial support came from the European network COSLAB.

\end{acknowledgments}


\begin{thebibliography}{10}

\bibitem{BayBlaHol99} G.\ Baym, J.-P.\ Blaizot, M.\ Holzmann, F.\ Lalo\"e, and D.\ Vautherin, Phys.\ Rev.\ Lett.\ {\bf 83}, 1703 (1999).

\bibitem{ArnMooTom01} P.\ Arnold, G.\ Moore, and B.\ Tom\'asik, Phys.\ Rev.\ A {\bf 65}, 013606 (2001).

\bibitem{KasProSvi01} V.~A.\ Kashurnikov, N.~V.\ Prokof'ev, and B.~V.\ Svistunov, Phys.\ Rev.\ Lett.\ {\bf 87}, 120402 (2001).

\bibitem{ArnMoo01} P.\ Arnold and G.\ Moore, Phys.\ Rev.\ Lett.\ {\bf 87}, 120401 (2001).

\bibitem{GioPitStr96}
S.\ Giorgini, L.~P.\ Pitaevskii, and S.\ Stringari, Phys.\ Rev.\ A {\bf 54},  R4633 (1996).

\bibitem{ArnTom01}
P.\ Arnold and B.\ Tom\'asik, Phys.\ Rev.\ A {\bf 64},  053609  (2001).

\bibitem{Zob04} O.\ Zobay, J.\ Phys.\ B {\bf 37}, 2593 (2004).

\bibitem{ZobMetAlb04} O.\ Zobay, G.\ Metikas, and G.\ Alber,
Phys.\ Rev.\ A {\bf 69}, 063615 (2004).

\bibitem{ShiZhe97a}
H.\ Shi and W.-M.\ Zheng, Phys.\ Rev.\ A {\bf 56}, 1046 (1997).

\bibitem{PI} H.\ Kleinert,
{\it Path Integrals in Quantum Mechanics, Statistics, Polymer Physics, and Financial Markets}, 3rd ed.\ (World Scientific, Singapore, 2004);
$\langle$http://www.physik.fu-berlin.de/{}\~{}{}kleinert/b8$\rangle$

\bibitem{MetZobAlb04} G.\ Metikas, O.\ Zobay, and G.\ Alber, Phys. Rev. A {\bf 69}, 043614 (2004).

\bibitem{SCT}
H.\ Kleinert, Phys.\ Rev.\ D {\bf 57}, 2264 (1998); {\bf 58}, 107702A (1998) (cond-mat/9803268);
Phys.\ Lett.\ B {\bf  434}, 74 (1998) (cond-mat/9801167).

\bibitem{KS} H.\ Kleinert and V.\ Schulte-Frohlinde, {\it Critical Properties of $\phi^4$-Theories} (World Scientific, Singapore, 2001).

\bibitem{SCT2}
H.\ Kleinert,
Phys.\ Rev.\ D {\bf 60}, 085001 (1999) (hep-th/9812197).

\bibitem{Lipa}
J.~A.\ Lipa, J.~A.\ Nissen, D.~A.\ Stricker, D.~R.\ Swanson, and T.~C.~P.\ Chui, Phys.\ Rev.\ B {\bf 68}, 174518 (2003).

\bibitem{Kle03} H.\ Kleinert, Mod.\ Phys.\ Lett.\ B {\bf 17}, 1011 (2003).

\bibitem{Kas03} B.\ Kastening, Phys.\ Rev.\ A {\bf 68}, 061601(R) (2003);
 {\bf 69}, 043613 (2004); {\bf 70}, 043621 (2004); Laser Phys. {\bf 14}, 586, (2004).

\bibitem{Zin89} J.\ Zinn-Justin, {\em Quantum Field Theory and Critical Phenomena} (Oxford University Press, Oxford, 1989).

\bibitem{FK}
R. P. Feynman
and H. Kleinert, Phys.\ Rev.\  A {\bf  34}, 5080 (1986).

\bibitem{GI}
{V.L. Ginzburg}, Fiz. Twerd. Tela (Leningrad) {\bf 2}, 2031 (1960)
[Sov. Phys. Solid State
{\bf 2}, 1824 (1961)];
see also the detailed discussion in Chapter 13 of the textbook
L.D. Landau and E.M. Lifshitz, {\em Statistical Physics\/},
3rd ed.\ (Pergamon, London, 1968).
Strictly speaking, the onset of fluctuations
in a complex U(1)-symmetric
field theory is not determined by Ginzburg's criterion
which concerns only size fluctuations, but by a modification of it
concerning the much more relevant
phase fluctuations; see
H.~Kleinert,
Phys.\ Rev.\ Lett.\ {\bf 84}, 286 (2000).

\bibitem{BagPriKle87} V.\ Bagnato, D.\ E.\ Pritchard, and D.\ Kleppner, Phys.\ Rev.\ A {\bf 35}, 4354 (1987).

\bibitem{LiCheChe99} M.\ Li, L.\ Chen, and C.\ Chen, Phys.\ Rev.\ A {\bf 59}, 3109 (1999).

\bibitem{Yan00} Z.\ Yan, Phys.\ Rev.\ A {\bf 61}, 063607 (2000).

\bibitem{ch7}
For detailed properties see Chapter 7
in
\cite{PI}.

\bibitem{HolBayBla01} M.\ Holzmann, G.\ Baym, J.-P.\ Blaizot, and F.\ Lalo\"e, Phys.\ Rev.\ Lett.\ {\bf 87}, 120403 (2001).

\bibitem{FetWal71}
A.~L.\ Fetter and J.~D.\ Walecka, {\it Quantum Theory of Many-Particle Systems} (McGraw-Hill, New York, 1971).

\bibitem{Rob51} J.~E.\ Robinson, Phys.\ Rev.\ {\bf 83}, 678 (1951).

\bibitem{KleSchPel04} H.\ Kleinert, S.\ Schmidt, and A.\ Pelster, Phys.\ Rev.\ Lett.\ {\bf 93}, 160402 (2004).

\end{thebibliography}
\end{document}